\documentclass[conference]{IEEEtran}%
\usepackage{amsfonts}
\usepackage{amsmath}
\usepackage{amssymb}
\usepackage{graphicx}%
\IEEEoverridecommandlockouts
\pdfoutput=1
\newtheorem{theorem}{Theorem}

\newtheorem{proposition}[theorem]{Proposition}

\let\originalleft\left
\let\originalright\right
\def\left#1{\mathopen{}\originalleft#1}
\def\right#1{\originalright#1\mathclose{}}
\begin{document}

\title{Polar coding to achieve the Holevo capacity of a pure-loss optical channel}
\author{\IEEEauthorblockN{Saikat Guha\thanks{\tiny{SG was supported by the DARPA Information in a Photon (InPho) program under DARPA/CMO Contract No. HR0011-10-C-0159. The views and conclusions contained in this document are those of the authors and should not be interpreted as representing the official policies, either expressly or implied, of the Defense Advanced Research Projects Agency or the U.S. Government. MMW acknowledges financial support from Centre de Recherches Math\'{e}matiques.}}}
\IEEEauthorblockA{\it Disruptive Information Proc. Tech. Group\\Raytheon BBN
Technologies\\Cambridge, Massachusetts, USA 02138}
\and \IEEEauthorblockN{Mark M. Wilde} \IEEEauthorblockA{\it School of Computer Science\\McGill University\\Montreal, Qu\'{e}bec,
Canada H3A 2A7}}

\maketitle

\begin{abstract}
In the low-energy high-energy-efficiency regime of classical optical communications---relevant to deep-space optical channels---there is a big gap between reliable communication rates achievable via conventional optical receivers and the ultimate (Holevo) capacity. Achieving the Holevo capacity requires not only optimal codes but also receivers that make collective measurements on long (modulated) codeword waveforms, and it is impossible to implement these collective measurements via symbol-by-symbol detection along with classical postprocessing~\cite{Gio04, Guh10a}. Here, we apply our recent results on the {\em classical-quantum polar code}~\cite{WG11}---the first near-explicit, linear, symmetric-Holevo-rate achieving code---to the lossy optical channel, and we show that it almost closes the entire gap to the Holevo capacity in the low photon number regime. In contrast, Arikan's original polar codes, applied to the DMC induced by the physical optical channel paired with any conceivable structured optical receiver (including optical homodyne, heterodyne, or direct-detection) fails to achieve the ultimate Holevo limit to channel capacity. However, our polar code construction (which uses the quantum fidelity as a channel parameter rather than the classical Bhattacharyya quantity to choose the ``good channels" in the polar-code construction), paired with a quantum successive-cancellation receiver---which involves a sequence of collective non-destructive binary projective measurements on the joint quantum state of the received codeword waveform---can attain the Holevo limit, and can hence in principle achieve higher rates than Arikan's polar code and decoder directly applied to the optical channel. However, even a theoretical recipe for construction of an optical realization of the quantum successive-cancellation receiver remains an open question. 
\end{abstract}

Determining the ultimate limits on optical communication must involve an explicitly quantum analysis, because electromagnetic waves are fundamentally quantum mechanical and high-sensitivity photodetection systems are limited by noise of quantum-mechanical origin. In quantum mechanics, the state of a physical system together with a description of the measurement made on that system determine the statistics of the measurement outcomes. Thus, in seeking the classical information capacity of an optical channel, we must allow for optimization over \em both\/\rm\ the transmitted quantum states \em and\/\rm\ the receiver's quantum measurement. In particular, it seems inappropriate to restrict consideration to coherent-state (laser) transmitters and coherent-detection or direct-detection receivers. Imposing these structural constraints leads to Gaussian-noise (Shannon-type) capacity formulas for coherent (homodyne and heterodyne) detection and Poisson-noise capacity results for shot-noise-limited direct detection~\cite{Gagliardi, MartinezPoisson}. None of these results, however, can be regarded as specifying the ultimate limit on reliable communication at optical frequencies. What is needed for deducing the fundamental limits on optical communication is an analog of Shannon's channel coding theorem---but free of unjustified structural constraints on the transmitter and receiver---that applies to data transmission over a {\em quantum channel}, viz., the Holevo-Schumacher-Westmoreland (HSW) theorem \cite{HSW1, Hau96, YO93}.

The HSW theorem, along with the Yuen-Ozawa converse \cite{YO93}, specifies the channel capacity of a pure-loss optical channel \cite{Gio04}. Even though the single-letter Holevo quantity is an achievable rate, the receiver in general must make joint-detection ({\em collective}) measurements over long codeword blocks---measurements that cannot be realized by detecting single-modulation symbols followed by classical decoding. For the pure-loss optical channel, a coherent-state modulation suffices to attain the ultimate capacity, i.e., use of non-classical transmitted states or entangled codewords does not increase capacity~\cite{Gio04}. We will use the term {\em Holevo capacity} unambiguously in this paper to refer to the single-letter Holevo rate of the pure-loss optical channel. The square-root-measurement, which in general is a positive operator-valued measure (POVM), applied to a random code gives the mathematical construct of a receiver measurement that can achieve the Holevo capacity \cite{Hau96}. The key questions that remain are how to design practical modulation formats, explicit codes (with efficient encoders), and most importantly, structured laboratory-realizable designs of Holevo-capacity-achieving joint-detection receivers (JDRs).

Lloyd {\em et al.}~\cite{Llo10} conceptualized a receiver that can attain the Holevo capacity of any quantum channel by making a sequence of ``yes/no" projective measurements on codewords of a random codebook. However, the translation of their strategy to a structured receiver design for the optical channel was not clear. Sen later simplified Lloyd {\em et al.}'s proof~\cite{S11}, and after this, we showed how to apply Sen's result in order to achieve the Holevo capacity of the pure-loss optical channel~\cite{GTW11}. The strategy employs a random code and a sequence of multi-mode phase-space displacements and quantum-non-demolition ``vacuum-or-not'' measurements. In Ref.~\cite{Guh10a}, one of us showed some of the first examples of structured optical JDRs for BPSK modulated binary codes, which induce superchannels over a codebook whose Shannon capacity per symbol is higher than the Shannon capacity of the single-use DMC induced by an optimal measurement on each received BPSK symbol~\cite{Dolinar}. In recent work, we generalized Arikan's {\em polar code} for the classical channel, to the {\em classical-quantum} (cq) polar code that can achieve the symmetric Holevo information for sending classical information over any quantum channel, i.e., the Holevo information rate when the input symbols are assigned equal priors~\cite{WG11}. This was the first explicit (and linear) code that provably achieves the Holevo capacity, and a careful inspection of \cite{WG11} reveals that this extension was non-trivial.

In this paper, we bring together our works from Refs.~\cite{Guh10a,WG11} to show that a cq-polar code essentially achieves the Holevo capacity of the optical channel with BPSK modulation. We show that at low mean pulse energy (photon number), this capacity is extremely close to the ultimate Holevo capacity (that with an unrestricted modulation). The receiver for our polar code is a quantum-limited successive-cancellation (SC) JDR that detects and decodes successively, while acting on the entire $N$-BPSK-symbol optical codeword waveform. It performs $NR$ binary-outcome, quantum-non-demolition measurements (as opposed to the $2^{NR}$ measurement stages in~\cite{Llo10, GTW11}), decoding one message bit at a time.
The final piece of the puzzle, a structured optical receiver design that implements our quantum SC decoder, remains a subject of ongoing research. 

\section{Capacity of the pure-loss optical channel}\label{sec:optical}

Consider a lossy optical channel with transmissivity $\eta \in (0, 1]$. Each channel use is a $T$-sec-long pulse slot that can transmit one modulation symbol. The mean energy\footnote{In this paper, we will use the term ``energy" to mean photon number. We are implicitly assuming a quasi-monochromatic light source with center frequency $\omega_0$, for which mean photon number is indeed proportional to energy (with a proportionality factor of ${\hbar\omega_0}$).} per transmitted pulse is constrained to $E$ photons per channel use. The Holevo capacity of this channel is given by $g({\eta}E)$ bits/use, where $g(x) = (1+x)\log(1+x) - x\log x$. A $\sqrt{{\rm photons/sec}}$-unit laser pulse $s(t)e^{j(\omega_0+\theta)t}, t \in [0, T)$ has energy $E_s = \int_0^T|s(t)|^2dt$ photons. Quantum-mechanically, the state of this pulse is a coherent state $|\alpha\rangle$, with $\alpha = \sqrt{E_s}e^{j\theta}$, where $\theta$ is taken w.r.t. some carrier-phase reference.\footnote{A general pure state of the temporal mode $s(t)$ is a unit-norm vector $|\psi\rangle = \sum_{n=0}^\infty c_n |n\rangle \in {\cal H}$, where $\left\{|n\rangle\right\}$ are quantum photon-number (Fock) states, which form a complete orthonormal basis of the Hilbert space $\cal H$. For a laser pulse, $c_n = \alpha^ne^{-|\alpha|^2/2}/\sqrt{n!}$ (photon number is Poisson-distributed with mean $|\alpha|^2$), and $|\psi\rangle$ in turn is the ``coherent state" $|\alpha\rangle$.} Since the channel preserves coherent states (with amplitude attenuation), $|\alpha\rangle \to |\sqrt{\eta}\alpha\rangle$, let us assume WLOG that $\eta = 1$ (or, equivalently, treat $E$ as the average received energy per pulse). The capacity-achieving input distribution is the isotropic Gaussian distribution $p(\alpha) = e^{-|\alpha|^2/E}/{\pi E}$, and the ultimate Holevo capacity, $C_{\rm ult}(E) = g(E)$ bits/symbol. 

Let us consider an equi-prior (received) BPSK alphabet $\left\{|\sqrt{E}\rangle, |-\sqrt{E}\rangle\right\}$. The minimum average probability of error in discriminating the two BPSK pulses is given by $P_{e, {\rm min}} = \left[1-\sqrt{1-e^{-4E}}\right]/2$, which can be attained exactly by the {\em Dolinar receiver}~\cite{Dolinar}. 
An ideal homodyne receiver achieves $P_{e, {\rm hom}} = {\rm erfc}\left(\sqrt{2E}\right)/2$. Using either the Dolinar receiver or the homodyne receiver on the BPSK pulses induces a binary symmetric channel (BSC) with crossover probabilities $P_{e, \min}$ and $P_{e, {\rm hom}}$, thus attaining channel capacities,\footnote{The subscript ``$1$" in $C_1(E)$ signifies that it is the highest capacity achievable using a single-symbol detection, for the BPSK alphabet.} $C_1(E) = 1 - H_2(P_{e,\min})$ and $C_{\rm hom}(E) = 1 - H_2(P_{e,{\rm hom}})$ bits/symbol, respectively, where $H_2(\cdot)$ is the binary entropy function. Another strategy is to use the Kennedy receiver, which coherently shifts the BPSK constellation to $\left\{|2\sqrt{E}\rangle, |\sqrt{0}\rangle\right\}$, followed by direct detection (Poisson statistics). This induces a Z-channel with crossover probability $e^{-4E}$ and with $P_{e, {\rm Ken}} = e^{-4E}/2$. 

\begin{figure}
\centering
\includegraphics[width=3.0in]{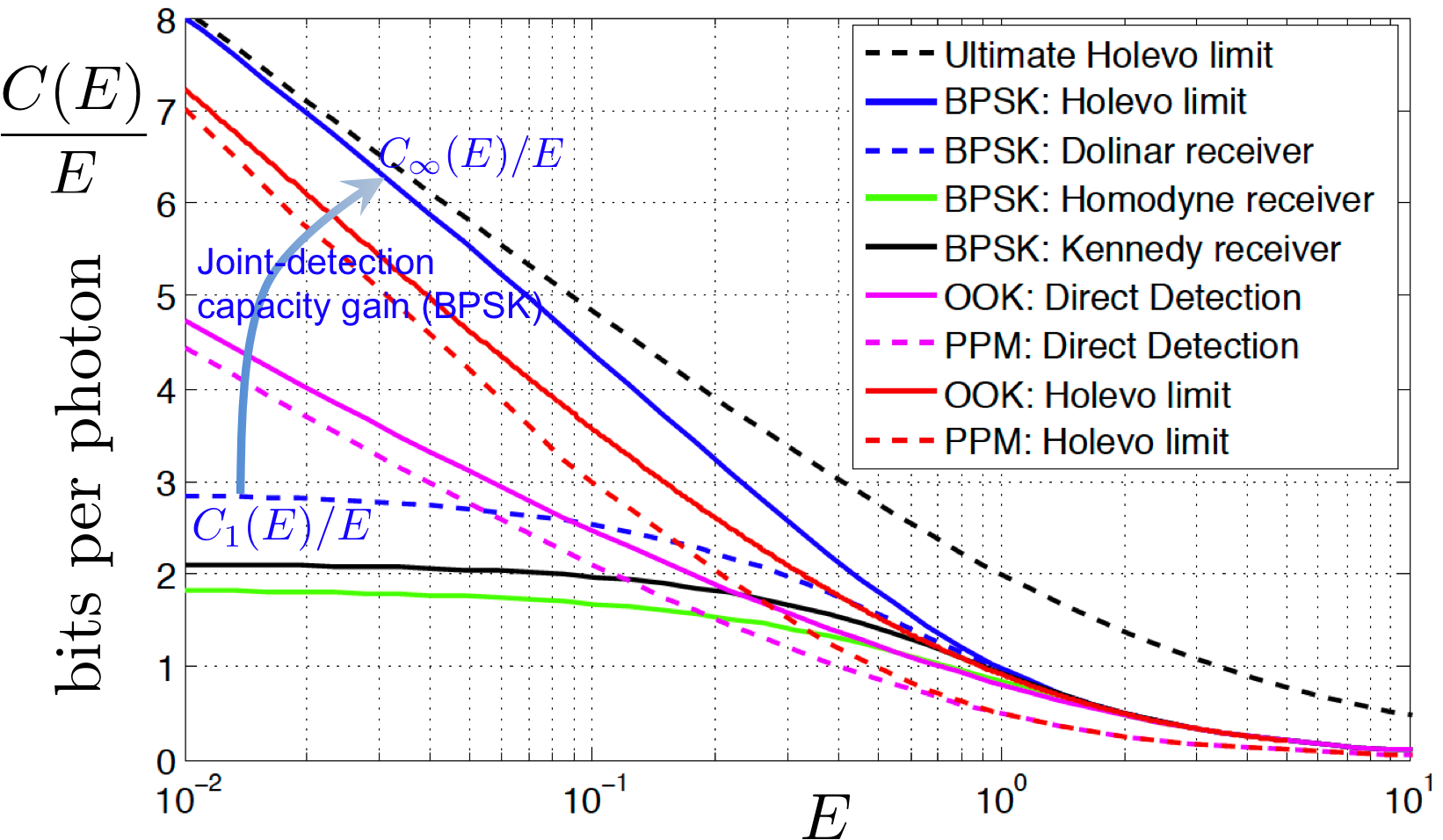}
\caption{Energy efficiency (bits/photon) vs. the mean pulse energy $E$. The arrow indicates the performance gap between Arikan's polar code paired with a classical decoder applied to the optical channel and our classical-quantum polar code paired with a quantum decoder.}
\label{fig:PIE}
\end{figure}
Consider a binary pure-state channel of the form $W:x\rightarrow\left\vert \psi_{x}\right\rangle$,
where $W$ denotes the channel, $x\in\left\{0,1\right\}$, and $\left\vert
\psi_{x}\right\rangle $ are pure-state channel outputs. For the BPSK channel, $|\psi_0\rangle = |\sqrt{E}\rangle$, and $|\psi_1\rangle = |-\sqrt{E}\rangle$. The relevant parameters that
determine channel performance are the fidelity
\begin{equation}
F\left(  W\right)  \equiv\left\vert \langle\psi_{0}|\psi_{1}\rangle\right\vert
^{2} = e^{-4E},
\label{eq:BPSKinnerproduct}
\end{equation}
and the symmetric Holevo information\footnote{The Holevo information reduces to the von Neumann entropy for a pure-state ensemble. More generally, if the channel outputs are mixed states $\rho_{0}$ and $\rho_{1}$, the fidelity is defined as $F\left(  W\right)  \equiv\left\Vert \sqrt{\rho_{0}}\sqrt{\rho_{1}}\right\Vert_{1}^{2},$
and the symmetric Holevo information as $I\left(  W\right)  \equiv H\left(  \left(  \rho_{0}+\rho_{1}\right)/2\right)  -H\left(  \rho_{0}\right)  /2-H\left(  \rho_{1}\right)  /2$.} $I\left(  W\right)  \equiv H\left(  \rho\right)$, with $\rho\equiv1/2\left(  \left\vert \psi_{0}\right\rangle \left\langle
\psi_{0}\right\vert +\left\vert \psi_{1}\right\rangle \left\langle \psi
_{1}\right\vert \right)$, where the von Neumann entropy $H\left(
\sigma\right)  \equiv-$Tr$\left\{  \sigma\log_{2}\sigma\right\}  $. The ultimate capacity achievable by the binary ensemble $|\psi_x\rangle$, $x \in \left\{0,1\right\}$, is given by the Holevo information of the average state $\rho_p \equiv p|\psi_0\rangle\langle{\psi_0}|+(1-p)|\psi_1\rangle\langle{\psi_1}|$ maximized over the prior $p \in [0, 1]$, i.e., $C = \max_{p}H\left({\rho_p}\right)$. The maximum for BPSK encoding is attained at $p = 1/2$. Therefore, the symmetric Holevo information is the Holevo capacity, $I(W) = H_2\left[\left({1 + \sqrt{F(W)}}\right)/{2}\right]$. Thus, for the BPSK alphabet, the Holevo capacity is given by:
\begin{equation}
C_\infty(E) = H_2\left[\left({1 + e^{-2E}}\right)/{2}\right],
\end{equation}
where the subscript ($\infty$) signifies that in order to achieve this capacity, the receiver must make collective measurements over long codeword blocks of an optimal binary code. 

It is well-known that for $E \gg 1$, a coherent-state modulation along with standard (symbol-by-symbol) heterodyne detection asymptotically achieves the ultimate Holevo capacity $g(E)$ bits/symbol~\cite{Gio04}, and that the capacity gap between conventional single-symbol receivers and the Holevo limit is the highest at low photon numbers ($E \ll 1$)~\cite{Guh10a}. Hence, for the rest of this paper, we will focus on this $E \ll 1$ regime. The energy efficiency $C_1(E)/E$ (bits/photon) of the BPSK channel, when the best single-symbol detection is used on each channel output symbol, caps off at $2$ nats/photon ($2.89$ bits/photon) as $E \to 0$. On the other hand, the Holevo limit to the energy efficiency of BPSK, $\lim_{E \to 0} C_{\infty}(E)/E = -\ln E + 1 + E\ln E + [\cdots]$ nats/photon~\cite{CGZ11} (where $[\cdots]$ indicates higher order terms), not only goes to infinity as $E \to 0$, but also approaches the ultimate (unrestricted-modulation) Holevo limit $g(E)/E$ asymptotically (see Fig.~\ref{fig:PIE}). The highest capacity (thus energy efficiency) with a BPSK modulation using a conventional receiver is achieved by ideal homodyne detection (see green plot in Fig.~\ref{fig:PIE}). Even though intensity modulation formats can attain an unbounded energy efficiency using a direct detection receiver, at low $E$ (we will come back to this in Section~\ref{sec:qary}), at $E \ll 1$, a BPSK code---along with a JDR---is capable of practically closing the gap all the way to the ultimate limit to capacity (which is not possible by an intensity-only modulation).

So, how do we understand this huge gap between the best single-symbol Shannon capacity and the Holevo capacity of the BPSK alphabet (gap shown by the arrow in Fig.~\ref{fig:PIE})? The two coherent states $\{|\sqrt{E}\rangle,|-\sqrt{E}\rangle\}$ are non-orthogonal (and thus not distinguishable), with inner product $e^{-2E} > 0$ as in \eqref{eq:BPSKinnerproduct}. However, by virtue of the HSW theorem~\cite{HSW1}, the joint quantum states of well-chosen (i.i.d. random) $2^{NR}$ sequences of these two states (codeword waveforms) become nearly perfectly distinguishable as $N \to \infty$ as long as $R < C_\infty(E)$. Since these codeword states live in the $N$-symbol Hilbert space ${\cal H}^{\otimes N}$, a collective measurement is required to discriminate these states at a vanishingly low error rate. If the best single-symbol detection is used to detect each output BPSK symbol (thereby inducing a BSC and \textit{we stress that this is the case with Arikan's polar encoding and classical successive cancellation decoder applied to the optical channel}), then an ML decoding can map the output {\em classical sequence} of these $N$ binary-outcome measurements to the correct codeword with a vanishingly low probability of error, however only as long as $R < C_1(E)$ (i.e., it can sustain a lower capacity). Classical information theory works with the classical-input to classical-output ``channel", which is determined by the combination of the physical transmission medium {\em and} the choice of the receiver measurement. Quantum information theory, in this case the HSW theorem, provides us with a tool to evaluate the best achievable capacity by automatically optimizing over all physically-realizable receiver measurements.

Unfortunately however, just like Shannon theory, the HSW theorem neither gives us a prescription to construct good low-complexity codes nor does it tell us how to realize the capacity-achieving receiver. In Section~\ref{sec:polarcode}, we provide the first explicit code (and a sequential-decoding collective measurement) that can achieve the BPSK Holevo capacity, the {\em classical-quantum polar code}.
\vspace{-5pt}
\section{Achieving the Holevo capacity of the binary pure-state quantum channel using a polar code}\label{sec:polarcode}

\label{sec:cq-review}We now demonstrate how to construct a polar code for the binary pure-state channel (a special case being the BPSK optical channel), by appealing to our recent results on cq-polar codes for sending classical data over a quantum channel. These codes achieve the symmetric Holevo information rate for a general (potentially mixed-state) binary input alphabet~\cite{WG11}.

\begin{figure}
\begin{center}
\includegraphics[
width=0.9\columnwidth
]
{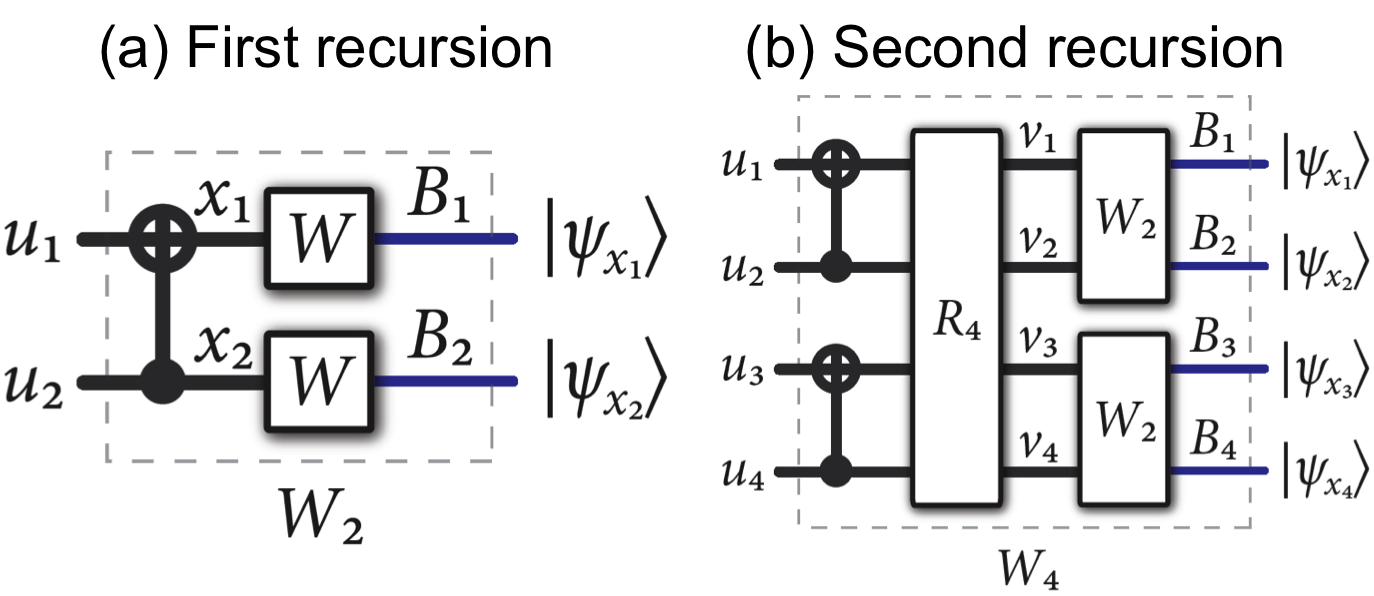}
\end{center}
\caption{(a) The channel $W_{2}$ synthesized from the first level of recursion.
Thick lines denote classical systems while thin lines quantum systems.
The gate shown is a CNOT, $\left(  u_{1},u_{2}\right)  \rightarrow\left(
u_{1}\oplus u_{2},u_{2}\right) $. (b) The second level of recursion in the channel combining phase. The operation $R_{4}$ simply permutes the bits so that all odd-index bits are
first and all even-index bits are last.}%
\label{fig:recursion}%
\end{figure}

Consider the binary pure-state classical-quantum channel $W:x\rightarrow\left\vert \psi_{x}\right\rangle$, $x\in\left\{0,1\right\}$. Channels with fidelity $F\left(  W\right)  \leq\epsilon$ are nearly noiseless and those with $F\left(  W\right)  \geq1-\epsilon$ are near to being completely useless. Recall that the quantum fidelity is a generalization of the classical Bhattacharya distance $Z$~\cite{A09,WG11}. Let us consider $N=2^{n}$ copies of $W$, such that the resulting channel is of the form:
$
x^{N}\equiv x_{1}\cdots x_{N}\rightarrow\left\vert \psi_{x^{N}}\right\rangle
\equiv\left\vert \psi_{x_{1}}\right\rangle \otimes\cdots\otimes\left\vert
\psi_{x_{N}}\right\rangle ,
$
where $x^{N}$ is the length $N$ input and $\left\vert \psi_{x^{N}%
}\right\rangle $ is the output state. We can extend Arikan's idea of channel
combining \cite{A09} to this classical-quantum channel, by considering the channels
induced by a transformation on an input bit (row) vector 
$
u^{N}\rightarrow|\psi_{u^{N}G_{N}}\rangle,
$
where $G_{N}=B_{N}F^{\otimes n}$, with $B_{N}$ being a permutation matrix that
reverses the order of the bits and 
$
F=%
\begin{bmatrix}
1 & 0\\
1 & 1
\end{bmatrix}
$.
This encoding is equivalent to a network of classical CNOT\ gates
and permutations that can be implemented with complexity $O\left(
N\log N\right)  $. See Fig.~\ref{fig:recursion}
for the first and second instances of this
encoding. Further instances are constructed recursively.
We then define \textquotedblleft split channels\textquotedblright\ from
the above combined channels as:%
\begin{equation}
W_{N}^{\left(  i\right)  }:u_{i}\rightarrow\rho_{\left(  i\right)  ,u_{i}%
}^{U_{1}^{i-1}B^{N}}, \label{eq:split-channels}%
\end{equation}
where,
\begin{align}
\rho_{\left(  i\right)  ,u_{i}}^{U_{1}^{i-1}B^{N}}  &  \equiv\sum_{u_{1}%
^{i-1}}\frac{1}{2^{i-1}}\left\vert u_{1}^{i-1}\right\rangle \left\langle
u_{1}^{i-1}\right\vert ^{U_{1}^{i-1}}\otimes\overline{\rho}_{u_{1}^{i}}%
^{B^{N}},\\
\overline{\rho}_{u_{1}^{i}}^{B^{N}}  &  \equiv\sum_{u_{i+1}^{N}}\frac
{1}{2^{N-i}}|\psi_{u^{N}G_{N}}\rangle\langle\psi_{u^{N}G_{N}}|^{B^{N}}.
\label{eq:averaged-cond-states}%
\end{align}
The interpretation of this channel is that it is the one \textquotedblleft
seen\textquotedblright\ by the bit $u_{i}$ if all the previous bits $u_{1}^{i-1}$ are available and if all the future bits $u_{i+1}^{N}$ as randomized. This motivates the development of a quantum
successive cancellation decoder \cite{WG11}\ that tries to distinguish $u_{i}=0$ from $u_{i}=1$ by adaptively exploiting the results of past measurements and Helstrom-Holevo measurements \cite{H69,Hol72}\ for each bit decision.

Arikan's polar coding rule divides the channels into \textquotedblleft
good\textquotedblright\ ones and \textquotedblleft bad\textquotedblright%
\ ones~\cite{A09}. Let $\left[  N\right]  \equiv\left\{  1,\ldots,N\right\}  $ and $0<\beta<1/2$. The polar coding rule for the
classical-quantum channel divides the channels as follows:%
\begin{align}
\mathcal{A} &  \equiv\left\{  i\in\left[  N\right]  :\sqrt{F(W_{N}^{\left(
i\right)  })}<2^{-N^{\beta}}\right\}  ,\label{eq:good-channels}
\end{align}
so that the channels in $\mathcal{A}$ are the good ones and those in
$\mathcal{A}^{c}$ are the bad ones. Observe that the quantum polar coding rule
involves the quantum fidelity parameter $F(\cdot)$, rather than a classical one such
as the Bhattacharya distance.

The following theorem is helpful in determining what fraction of the channels
become good or bad \cite{AT09}:

\begin{theorem}
[Convergence Rate]\label{thm:conv-rate}Let $\left\{  X_{n}:n\geq0\right\}  $
be a random process with $0\leq X_{n}\leq1$ and satisfying%
\begin{align}
X_{n+1}  &  \leq qX_{n}\ \ \ \ \ \text{w.p.\ \ \ }%
1/2,\label{eq:general-process-relations-4-conv}\\
X_{n+1}  &  =X_{n}^{2}\ \ \ \ \ \text{w.p.\ \ \ }1/2,
\label{eq:general-process-relations-4-conv-2}%
\end{align}
where $q$ is some positive constant. Let $X_{\infty}=\lim_{n\rightarrow\infty
}X_{n}$ exist almost surely with $\Pr\left\{  X_{\infty}=0\right\}
=P_{\infty}$. Then for any $\beta<1/2$,
$
\lim_{n\rightarrow\infty}\Pr\{  X_{n}<2^{-2^{n\beta}}\}
=P_{\infty},
$
and for any $\beta>1/2$,
$
\lim_{n\rightarrow\infty}\Pr\{  X_{n}>2^{-2^{n\beta}}\}  =0.
$
\end{theorem}

The channel combining and splitting mentioned above can be considered as a
random birth process in which a channel $W_{n+1}$ is constructed from two
copies of a previous one $W_{n}$ according to the rules in Section~4 of
Ref.~\cite{WG11}. One can then consider the process $\left\{  F_{n}%
:n\geq0\right\}  \equiv\{  \sqrt{F\left(  W_{n}\right)  }:n\geq0\}
$ and prove that it is a bounded super-martingale by exploiting the
relationships given in Proposition~10 of Ref.~\cite{WG11}. From the
convergence properties of martingales, it follows that $F_{\infty}$
converges almost surely to a value in $\left\{  0,1\right\}  $, and the
probability that it equals zero is equal to the symmetric Holevo
information$~I\left(  W\right)  $. Furthermore, since the process $F_{n}$
satisfies the relations in (\ref{eq:general-process-relations-4-conv}%
-\ref{eq:general-process-relations-4-conv-2}), the following proposition on
the convergence rate of polarization holds:

\begin{theorem}
\label{thm:fraction-good}Given a binary input classical-quantum channel $W$
and any $\beta<1/2$,
$
\lim_{n\rightarrow\infty}\Pr\{  F_{n}<2^{-2^{n\beta}}\}  =I\left(
W\right)
$.

\end{theorem}

One of our important advances in Ref.~\cite{WG11} was to establish that a
quantum successive cancellation decoder performs well for polar coding over
classical-quantum channels with equiprobable inputs. Corresponding to the split channels $W_{N}%
^{\left(  i\right)  }$ in (\ref{eq:split-channels}) are the following
projectors that attempt to decide whether the input of the $i^{\text{th}}$
split channel is zero or one:%
\begin{align*}
\Pi_{\left(  i\right)  ,0}^{U_{1}^{i-1}B^{N}} &  \equiv\left\{  \rho_{\left(
i\right)  ,0}^{U_{1}^{i-1}B^{N}}-\rho_{\left(  i\right)  ,1}^{U_{1}^{i-1}%
B^{N}}\geq0\right\}  ,\\
\Pi_{\left(  i\right)  ,1}^{U_{1}^{i-1}B^{N}} &  \equiv I-\Pi_{\left(
i\right)  ,0}^{U_{1}^{i-1}B^{N}},
\end{align*}
where $\left\{  B\geq0\right\}  $ denotes the projector onto the positive
eigenspace of a Hermitian operator~$B$. After some calculations,
we can readily see that%
\begin{align}
\Pi_{\left(  i\right)  ,0}^{U_{1}^{i-1}B^{N}} &  =\sum_{u_{1}^{i-1}}\left\vert
u_{1}^{i-1}\right\rangle \left\langle u_{1}^{i-1}\right\vert ^{U_{1}^{i-1}%
}\otimes\Pi_{\left(  i\right)  ,u_{1}^{i-1}0}^{B^{N}}%
,\label{eq:projectors-expanded-1}
\end{align}
where $\Pi_{\left(  i\right)  ,1}^{U_{1}^{i-1}B^{N}} = I - \Pi_{\left(  i\right)  ,0}^{U_{1}^{i-1}B^{N}}$,
$\Pi_{\left(  i\right)  ,u_{1}^{i-1}0}^{B^{N}}   \equiv \{  \overline
{\rho}_{u_{1}^{i-1}0}^{B^{N}}-\overline{\rho}_{u_{1}^{i-1}1}^{B^{N}}%
\geq0\}$  ,
$\Pi_{\left(  i\right)  ,u_{1}^{i-1}1}^{B^{N}}   \equiv I - \Pi_{\left(  i\right)  ,u_{1}^{i-1}0}^{B^{N}}
$.

The above observations lead to a decoding rule for a successive cancellation decoder
similar to Arikan's~\cite{A09}:%
\[
\hat{u}_{i}=\left\{
\begin{array}
[c]{cc}%
u_{i} & \text{if }i\in\mathcal{A}^{c}\\
h\left(  \hat{u}_{1}^{i-1}\right)   & \text{if }i\in\mathcal{A}%
\end{array}
\right.  ,
\]
where $h\left(  \hat{u}_{1}^{i-1}\right)  $ is the outcome of the following
$i^{\text{th}}$ (collective) measurement on the codeword received at the channel output (after $i-1$
measurements have already been performed):
$
\{  \Pi_{\left(  i\right)  ,\hat{u}_{1}^{i-1}0}^{B^{N}},\Pi_{\left(
i\right)  ,\hat{u}_{1}^{i-1}1}^{B^{N}}\}  .
$
We are assuming that the measurement device outputs \textquotedblleft%
0\textquotedblright\ if the outcome $\Pi_{\left(  i\right)  ,\hat{u}_{1}%
^{i-1}0}^{B^{N}}$ occurs and it outputs \textquotedblleft1\textquotedblright%
\ otherwise. (Note that we can set $\Pi_{\left(  i\right)  ,\hat{u}_{1}%
^{i-1}u_{i}}^{B^{N}}=I$ if the bit $u_{i}$ is a frozen bit.) The above
sequence of measurements for the whole bit stream $u^{N}$ corresponds to a
positive operator-valued measure (POVM)~$\left\{  \Lambda_{u^{N}}\right\}  $
where%
\begin{multline*}
\Lambda_{u^{N}}\equiv\Pi_{\left(  1\right)  ,u_{1}}^{B^{N}}\cdots\Pi_{\left(
i\right)  ,u_{1}^{i-1}u_{i}}^{B^{N}}\cdots\\
\cdots\Pi_{\left(  N\right)  ,u_{1}^{N-1}u_{N}}^{B^{N}}\cdots\Pi_{\left(
i\right)  ,u_{1}^{i-1}u_{i}}^{B^{N}}\cdots\Pi_{\left(  1\right)  ,u_{1}%
}^{B^{N}},
\end{multline*}
and $\sum_{u_{\mathcal{A}}}\Lambda_{u^{N}}=I^{B^{N}}$.

The probability of error $P_{e}\left(  N,K,\mathcal{A},u_{\mathcal{A}^{c}%
}\right)  $\ for code length $N$, number $K$ of information bits, set
$\mathcal{A}$ of information bits, and choice $u_{\mathcal{A}^{c}}$ for the
frozen bits is as follows:%
\[
P_{e}\left(  N,K,\mathcal{A},u_{\mathcal{A}^{c}}\right)  =1-\frac{1}{2^{K}%
}\sum_{u_{\mathcal{A}}}\text{Tr}\left\{  \Lambda_{u^{N}}\rho_{u^{N}}\right\}
,
\]
where we are assuming a particular choice of the bits $u_{\mathcal{A}^{c}}$ in
the sequence of projectors $\Pi_{\left(  N\right)  ,u_{1}^{N-1}u_{N}}^{B^{N}}$
$\cdots$ $\Pi_{\left(  i\right)  ,u_{1}^{i-1}u_{i}}^{B^{N}}$ $\cdots$
$\Pi_{\left(  1\right)  ,u_{1}}^{B^{N}}$ and $\Pi_{\left(  i\right)  ,u_{1}^{i-1}u_{i}}^{B^{N}}=I$ if $u_{i}$ is a
frozen bit. We are also assuming that the sender transmits the information
sequence $u_{\mathcal{A}}$ with uniform probability $2^{-K}$. The probability
of error averaged over all choices of the frozen bits is then,
\begin{align}
&P_{e}\left(  N,K,\mathcal{A}\right) =\frac{1}{2^{N-K}}\sum_{u_{\mathcal{A}^{c}}}P_{e}\left(  N,K,\mathcal{A},u_{\mathcal{A}^{c}}\right)   \nonumber
\end{align}

The following proposition from Ref.~\cite{WG11} determines the average
ensemble performance of polar codes with a quantum successive cancellation decoder:
\begin{proposition}
\label{prop:error-bound}For any classical-quantum channel $W$ with binary
inputs and quantum outputs and any choice of $\left(  N,K,\mathcal{A}\right)
$, the following bound holds%
\[
P_{e}\left(  N,K,\mathcal{A}\right)  \leq2\sqrt{\sum_{i\in\mathcal{A}}\frac
{1}{2}\sqrt{F(W_{N}^{\left(  i\right)  })}}.
\]
\end{proposition}

We proved the above bound on the performance of our SC decoder by exploiting
Sen's \textquotedblleft non-commutative union bound\textquotedblright%
~\cite{S11} and Lemma 3.2 of Ref.~\cite{H06} (which upper bounds the probability of error in
a binary quantum hypothesis test by the fidelity between the test states). The bound holds under the assumption that the sender chooses the information bits $U_{\mathcal{A}}$ from a uniform distribution. Thus, by choosing the channels over which the sender transmits the information
bits to be in $\mathcal{A}$ and those over which she transmits agreed-upon
frozen bits to be in $\mathcal{A}^{c}$, we obtain the following bound on the
probability of decoding error, as long as the code rate $R = K/N < I(W)$:
$
\Pr\{\widehat{U}_{\mathcal{A}}\neq U_{\mathcal{A}}\}=o(2^{-\frac{1}{2}%
N^{\beta}}).
$
This completes the specification of a cq polar code.

\subsubsection{Polar codes for $q$-ary input channels}\label{sec:qary}

The binary {\em on-off keying} (OOK) alphabet $\left\{|0\rangle, |\sqrt{E_0}\rangle\right\}$ with priors $(1-p^*, p^*)$, $E = p^*E_0$, with optimal $p^*(E) \approx -E\ln E/3$, $E \ll 1$, along with a (symbol-by-symbol) direct detection (DD) receiver, attains a photon efficiency $C_{\rm OOK-DD}(E)/E = -\ln E - \ln\ln(1/E) + [{\rm \cdots}]$ nats/photon~\cite{CGZ11} (magenta solid plot in Fig.~\ref{fig:PIE}). Furthermore, the Holevo capacity of the OOK alphabet is attained by $p^*(E) \approx \sqrt{E/2}$, and the photon efficiency is given by $C_{\rm OOK-Holevo}(E)/E = -\ln E + 1 + \sqrt{2}E^{1/2}\ln E + [{\rm \cdots}]$ nats/photon (solid red plot in Fig.~\ref{fig:PIE}). At $E \ll 1$, a $q$-ary PPM constellation (which can be seen as a rate-$(\log_2q)/q$ code over an underlying OOK alphabet) achieves a Shannon capacity (with DD) and a Holevo capacity, which are both extremely close to the respective unrestricted OOK capacities (dashed magenta and red plots in Fig.~\ref{fig:PIE}) respectively.

The $q$-ary PPM constellation achieves its capacity (both DD-Shannon and Holevo) for a uniform prior over its $q$ inputs, forming a $q$-ary input classical-quantum channel. If $q$ is a power of two, then a polar coding strategy to achieve the Holevo limit of PPM is straightforward, following the strategy to polar code for a uniform-input $q$-ary classical DMC (which can in turn achieve the PPM-DD capacity)~\cite{STA09}. Suppose that $m\equiv\log_2(q)$. Then one can decompose the input variable $X$ as an $m$-fold Cartesian product of binary variables $(X_1, \ldots, X_m)$ and exploit a polar code for each of these variables. One first exploits a quantum successive cancellation (SC) decoder to decode the variable $X_1$ under the assumption that the other variables $X_2$, \ldots, $X_m$ are chosen uniformly at random (and \textit{thus are independent}) for this first step. This decoding achieves a low probability of error as long as the indices for the information bits are chosen according to the polar coding rule for this first induced channel. After decoding $X_1$, the quantum measurement could potentially disturb the state at the channel output, but this disturbance will be asymptotically small if the measurement successfully decodes (a result known as the Gentle Measurement Lemma \cite{itit1999winter}). Then, $X_1$ is available as side information for decoding the next variable $X_2$, and the procedure extends iteratively by decoding the current variable $X_i$ with the previous $i$ ones available as side information and randomizing over the future $m-i$ variables. The rate achieved with this scheme is equal to the symmetric Holevo capacity, by exploiting the chain rule and independence:
$
I(X_1 \cdots X_m; B) = \sum_{i=1}^m I(X_i; B X^{i-1}_1).
$

\section{Discussions and Conclusion}

Our polar code and decoder construction in Ref.~\cite{WG11} offers the first near-explicit construction that almost closes the gap to the Holevo capacity limit for low-photon-number (high photon-efficiency) optical communications. Our construction improves upon earlier schemes by providing an explicit linear code with an efficient encoder (as opposed to a random code), while exponentially reducing the number of decoding steps ($NR$ steps as opposed to the $2^{NR}$ in Refs.~\cite{Llo10, GTW11}). 

Several practical questions remain unanswered, the most important one perhaps being an explicit design of our polar-decoding receiver, i.e., an optical circuit involving beamsplitters, phase-shifters, squeezers, and potentially one third-order Hamiltonian such as a Kerr interaction. In order to make this scheme practical, finding efficient means to compute the rate matched cq polar codes for quantum channels would be necessary. Finally, it would be interesting to find an efficient classical-quantum polar coding scheme that can handle non-uniform input priors (viz., to achieve the Holevo limit of the OOK modulation alphabet).



\bibliographystyle{IEEEtran}
\bibliography{Ref}

\end{document}